\documentclass[aps,preprint,showpacs,preprintnumbers,amsmath,amssymb,floatfix]{revtex4}
\usepackage[german,english]{babel}
\usepackage{bm,amsmath,amsfonts}
\usepackage{amssymb}
\usepackage{graphicx}
\usepackage{xcolor}

\def\beq{\begin{equation}}
\def\eeq{\end{equation}}
\def\beqn{\begin{eqnarray}}
\def\eeqn{\end{eqnarray}}

\def\ul {\underline}

\begin{document}

\title{Exact quantum dynamics of bosons with finite-range time-dependent interactions of harmonic type}

\author{Axel U. J. Lode$^{1}$, Kaspar Sakmann$^{1}$,
Ofir E. Alon$^{2}$, Lorenz S. Cederbaum$^{1}$, and Alexej I. Streltsov$^{1}$\footnote{E-mail: Alexej.Streltsov@pci.uni-heidelberg.de} }

\affiliation{$^1$ Theoretische Chemie, Physikalisch-Chemisches Institut, Universit\"at Heidelberg,\\
Im Neuenheimer Feld 229, D-69120 Heidelberg, Germany}

\affiliation{$^2$ Department of Physics, University of Haifa at Oranim, Tivon 36006, Israel}

\begin{abstract}
The exactly solvable quantum many-particle model with 
harmonic one- and two-particle interaction terms is extended to include time-dependency.
We show that when the external trap potential and finite-range interparticle interaction
have a time-dependency the exact solutions
of the corresponding time-dependent many-boson Schr\"odinger equation are still available.
We use these exact solutions to benchmark the recently developed multiconfigurational
time-dependent Hartree method for bosons (MCTDHB)
[Phys. Rev. Lett. {\bf 99}, 030402 (2007), Phys. Rev. A {\bf 77}, 033613 (2008)].
In particular, we benchmark the MCTDHB method for:
(i) the ground state;
(ii) the breathing many-body dynamics activated by a quench scenario 
where the interparticle interaction strength is suddenly turned on to a finite value;
(iii) the non-equilibrium dynamic for driven scenarios where 
both the trap- and interparticle-interaction potentials are {\it time-dependent}.
Excellent convergence of the ground state and dynamics is demonstrated.
The great relevance of the self-consistency and time-adaptivity,
which are the intrinsic features of the MCTDHB method, is demonstrated by
contrasting the MCTDHB predictions and those obtained 
within the standard full configuration interaction method
spanning the Fock space of the same size,
but utilizing as one-particle basis set the fixed-shape eigenstates of the one-particle potential.
Connections of the model's results to ultra-cold Bose-Einstein condensed systems are addressed.
\end{abstract}

\maketitle

\section{Introduction}

Since the first realizations of Bose-Einstein condensates (BECs) \cite{BECketterle,BECwieman,bradley:95}
the experiments on this unique state of quantum systems have become more and more complex.
Nowadays, the strength of the interparticle interactions, the trapping potential
and the dimensionality of BECs are under experimental control \cite{1d1,1d2,1d3,Feshbach_review,Henderson:09}.
This makes BECs a vivid and rich testing ground for a wide range of physical theories.
Recent realizations of the dipolar BECs \cite{Cr_exp_1,Cr_exp_2,Dy_exp_1} open a new perspective 
in the development of the physics of ultra-cold atoms and molecules.
A control on a new degree-of-freedom is achieved
 -- the dipolar long-range part of the interparticle interaction can be now customized.
This achievement can be considered as a first successful step towards a control on the overall shape of the interparticle interaction.
It also stimulates the development of theoretical methods capable to
solve the time-(in)dependent many-particle Schr\"odinger equation (TDSE) which
governs the physics of the trapped ultra-cold systems with general interparticle interactions.
The class of the many-body Hamiltonians which permits analytical solutions is quite small,
so, generally, one has to rely on numerical many-body methods to solve the TDSE.
The many-body methods in use have to be qualified to describe quantum many-body statics and dynamics.
Benchmarking of these methods against exactly-solvable Hamiltonians is a necessary step
for such a qualification.

In this work we consider an exactly solvable many-body Hamiltonian, where both the one-body (trap)
and two-body (interparticle interaction) potentials are of harmonic type,
also known as the harmonic interaction model (HIM), see Refs.~\cite{Cohen:85,Yan:03}.
The exact solutions of the HIM problem are obtained by transformation
of the Hamiltonian from the laboratory to the center of mass frame, where the Hamiltonian becomes separable.
The price of the transformation is that an intuitive physical picture of ``real'' particles is lost
and, instead, one operates with effective ``particles'' representing the transformed coordinates.
One wants to have, first, a general many-body method for identical physical particles
where each particle has its own ``real'' coordinate.
Second, the method must be powerful enough to solve problems where such ``real'' coordinates are not favorable (suitable).
Third, it should be capable to solve general problems where separations (transformations) of the variables are impossible,
as it is the case in unharmonic and multiwell traps.
In this work we want to test such a method -- the recently developed multiconfigurational time-dependent Hartree for bosons
(MCTDHB) method \cite{streltsov:07,alon:08},
which can treat the dynamics, i.e., the TDSE for trapped bosons with general interparticle interactions. 
We would like to carry the examples to an extreme case separable in suitable coordinates;
in this way we can unambiguously test the performance of the MCTDHB against analytical/exact solutions.

While there are several known many-body models with time-independent Hamiltonians which have analytical solutions,
the exactly solvable time-dependent many-body Hamiltonians are even less abundant. Why one needs to study them?
Apart from an exploration of novel dynamical physical phenomena, there is a practical reason for it.
In typical experiments with ultra-cold systems the manipulations of the trapping potentials,
as well as altering of the magnetic field used in the Feshbach resonance technique(s) \cite{Feshbach_review}
to manipulate the interparticle interaction, are time-dependent procedures.
So, there is a need for a proven theoretical method capable
to solve time-dependent Hamiltonians where both the trap and interparticle interaction potentials are time-dependent.
However, the TDSE with general time-dependent Hamiltonians can be solved only numerically,
hence it is very difficult to verify and quantify the region of applicability and quality of the numerical solutions obtained.
Convincing comparisons/benchmarks against exact results are of great relevance.
In this work we show how to extend the exactly solvable quantum many-particle HIM problem to include time-dependency,
and use it to benchmark the MCTDHB method.

It is worthwile to mention that some physical phenomena and properties of the many-body solutions 
of the HIM problem are ``universal'', i.e.,
transferable to systems with other interparticle interactions, e.g., contact interaction.
For example, small displacements of the density out of the center of an harmonic trap
result in so called ``dipole oscillations'' with the trap frequency which are independent of the interparticle interaction.
Another example is a quench of the interparticle interaction in an harmonically trapped system --
it activates only ``breathing`` excitations which preserve the symmetry of the trap.
In the present work we discover a novel time-dependent phenomenon in the extended HIM,
and discuss its ``universality'' for the harmonically trapped systems with general interparticle interactions.

The structure of the paper is as follows.
In Sec.~\ref{HIMdfns} we introduce the harmonic interaction model and discuss the aftermaths and implications
appearing due to the transformation of the coordinates from the center of mass frame,
where the exact solutions are analytically known, to the laboratory frame where we want to solve the problem numerically.
The MCTDHB method is briefly reviewed in Sec.~\ref{MCTDHB_th}.
Sec.~\ref{GS} provides detailed benchmarks and comparisons of the exact and numerical results for the ground state of the HIM
obtained within the framework of the MCTDHB and the standard full configuration interaction (FCI) methods,
the latter is also known as the exact diagonalization (ED) technique.
In Sec.~\ref{HIMquench} we benchmark our numerical tools to describe the breathing many-body dynamics
activated by a quench scenario where the interparticle interaction strength is suddenly
turned on from zero to a finite value.
Sec.~\ref{TDHIM} shows how to extend the exactly solvable quantum many-particle model with 
harmonic one- and two-particle interaction terms to include time-dependency.
Here we also demonstrate the applicability of the MCTDHB method to describe numerically-exact many-boson dynamics
for complicated scenarios where both the external trap and interparticle interaction potentials are time-dependent.
Sec.~\ref{sum} summarizes our results and outlooks the novel predictions obtained for the HIM problem to
ultra-cold atomic systems with contact interactions.

\section{The Harmonic Interaction Model (HIM)}\label{HIMdfns}
\subsection{Basic Definitions}\label{HIMdf_A}

Our starting point is the harmonic interaction model (HIM), see, e.g., Refs.~\cite{Cohen:85,Yan:03}.
The Hamiltonian of the HIM is readily obtained in the laboratory frame of reference
by setting the interparticle interaction potential $\hat{W}$
and the one-body potential $\hat{V}$ in the many-body Hamiltonian in dimensionless units, 
\begin{equation}
\hat{H}=\sum_{i=1}^N \left( \hat{T}(\vec{r}_i) + \hat{V}(\vec{r}_i) \right) + \sum_{i<j}^{N} \hat{W}(\vec{r}_i,\vec{r}_j),
\label{HAM}
\end{equation}
to be harmonic:
\begin{equation}
\hat{W}(\vec{r}_i,\vec{r}_j)= K_0 \left( \vec{r}_i - \vec{r}_j \right)^2,\qquad \hat{V}(\vec{r}) = \frac12 \omega^2 \vec{r}^2
\label{HIMint}.
\end{equation}
Here, $K_0$ accounts for the strength of the two-body interaction and $\hat{T}(\vec{r}) = - \frac{1}{2}\partial_{\vec{r}}^2$ is the kinetic energy operator. A positive value of $K_0$ corresponds to an attraction while a negative value means repulsion. In the case of a parabolic trapping potential, it is easy to see that the system becomes unbound when the value of $K_0$ is negative and big enough for the two-body repulsion to overcome the one-body harmonic trapping, i.e. $K_0< - \frac{\omega^2}{2 N}$.

Following Cohen and Lee in Ref.~\cite{Cohen:85}, the Hamiltonian, Eqs.~(\ref{HAM},\ref{HIMint}),
can be separated into $N$ independent harmonic oscillators by the following coordinate transformations:
\begin{eqnarray}
\vec{q}_j&=&\frac{1}{\sqrt{j(j+1)}} 
\sum_{i=1}^{j} (\vec{r}_{j+1} - \vec{r}_i), \qquad j=1,...,N-1 ,\\ \nonumber
\vec{q}_N&=&\frac{1}{\sqrt{N}}\sum_{i=1}^N \vec{r}_i. \label{coordtrans}
\end{eqnarray}
The transformed Hamiltonian in the center of mass frame reads:
\begin{eqnarray}
\hat{H} &=& \hat{H}_{rel} + \hat{H}_{CM}, \nonumber \\
\hat{H}_{rel}&=& \sum_{i=1}^{N-1} (- \frac12 \partial^2_{\vec{q}_i}+ \frac12 \delta^2_N \vec{q}^2_i),\qquad \label{HIMTD} \\ 
\qquad \hat{H}_{CM}&=& - \frac12 \partial^2_{\vec{q}_N} + \frac12 \omega^2 \vec{q}^2_N. \nonumber
\end{eqnarray}
Here, $\delta_N=\sqrt{\omega^2 + 2 N K_0}$ is the trapping frequency of the $N-1$ harmonic oscillators 
originating from the set of relative coordinates;
and $1$ harmonic oscillator with the frequency $\omega$ representing a center of mass coordinate. 
This separability of the HIM Hamiltonian into the center of mass and relative coordinates allows the following visualization:
the overall HIM system can be pictured as a medium formed by $N-1$ identical, noninteracting particles 
associated with relative coordinates $q_k,k=1,...,N-1$, 
moving in an effective harmonic trap with a time-independent frequency $\delta_N$,
and an independent effective particle with coordinate $q_N$, representing the system's center of mass,
trapped in the original time-independent harmonic potential with frequency $\omega$.

The general solution of the HIM problem in its separable form, Eq.~(\ref{HIMTD}), is
a product of $N$ generally different harmonic oscillator wavefunctions, and 
the total energy is the sum of the corresponding oscillator's energies.
The exact energy $E_{exact}$ of the ground state takes on a very simple form, see, e.g., Refs.~\cite{Cohen:85,Yan:03}:
\begin{eqnarray}
E_{exact}&=&\frac{D}{2} (N-1) \delta_N + \frac{D}{2}\omega.
\label{himEexact}
\end{eqnarray}
Here, $D$ is the dimensionality of the HIM system. We note that the HIM problem 
is an example of a many-body system with finite-range interparticle interactions
which permits analytical solution in any dimension, i.e., in 1D and in higher dimensions.
This is another attractive feature of the HIM model relevant for benchmarking numerical methods for the many-particle TDSE. 

\subsection{Representing the HIM with Basis Functions}

Let us consider the ground state of the (one-dimensional) HIM problem for $N=2$ bosons.
We contrast the solution written in the center of mass frame, i.e., in the $q_1, q_2$ coordinates
with that in the laboratory frame, i.e., in $x_1, x_2$ coordinates. The corresponding 
transformations of the coordinates are given in Eq.~(\ref{coordtrans}) with $\vec{q}_j=q_j$ and $\vec{r}_j=x_j$.
It is convenient to denote the $n$-th harmonic oscillator (HO) function as $\psi^{ho}_n(X,\Omega)$, here 
$\Omega$ is harmonic oscillator frequency and $X$ a general variable.
The ground state solution of the two-particle HIM problem reads:
\begin{eqnarray}
\Psi(q_1,q_2)&=&\psi^{ho}_0(q_1,\delta_2)\psi^{ho}_0(q_2,\omega)=
\mathcal{N}  e^{-\frac12 \delta_2 q^2_1}  e^{-\frac12 \omega q^2_2} \equiv \nonumber \\  \nonumber
\Psi(x_1,x_2)&=& 
\mathcal{N}  e^{-\frac12 \delta_2 (\frac{1}{\sqrt{2}}(x_2-x_1))^2} e^{-\frac12 \omega (\frac{1}{\sqrt{2}}(x_2+x_1))^2} \nonumber \\
&=&
\mathcal{N} e^{-\frac{\delta_2+\omega}{4}x_2^2} e^{-\frac{\delta_2+\omega}{4}x_1^2}e^{-\frac12 (\omega-\delta_2) x_2 x_1}  \nonumber \\
  &=& \mathcal{N}\psi^{ho}_0(x_1,\frac{\omega+\delta_2}{2})\psi^{ho}_0(x_2,\frac{\omega+\delta_2}{2}) e^{-\frac12 (\omega-\delta_2) x_2 x_1}  \nonumber \\
  &=& \mathcal{N}\psi^{ho}_0(x_1,\frac{\omega+\delta_2}{2})\psi^{ho}_0(x_2,\frac{\omega+\delta_2}{2})
 \left( 1 -\frac12 (\omega-\delta_2) x_2 x_1 + \frac{1}{8}(\omega-\delta_2)^2 x^2_2 x^2_1 - \ldots \right) \nonumber \\
&=& \sum_{i\ge j=0}^{\infty} a_{ij} \hat{\mathcal{S}} \psi^{ho}_i(x_1,\frac{\omega+\delta_2}{2})\psi^{ho}_j(x_2,\frac{\omega+\delta_2}{2}).
\label{HIMGS2bosons}
\end{eqnarray}
Here we use the Taylor expansion for the cross-term $e^{-\frac12 (\omega-\delta_2) x_2 x_1}$.
$\hat{\mathcal{S}}$ is the symmetrization operator and $a_{ij}$ -- known reexpansion coefficients.
The close inspection of the above transformation shows that a single Hartree product of two HO wavefunctions written 
in the center of mass frame is represented by an infinite sum of different Hartree products in the laboratory frame
even if one uses the ``dressed'' frequencies $\frac{\omega+\delta_2}{2}$. 
For bosonic systems these Hartree products have to be properly symmetrized to take into account permutational symmetry
of the total wavefunction.
Hence, a numerical solution of the HIM Hamiltonian in the laboratory frame is a very involved problem --
the numerical convergence depends on how fast and efficient this sum is spanned.

For numerical treatments the infinite sums, like in (\ref{HIMGS2bosons}), must be truncated.
The number of the terms $N_\mathrm{conf}$ considered defines the size of the Fock space spanned,
i.e., the size of the respective secular matrix to be diagonalized in order to find the respective eigenvalues and eigenstates.
For a general $N$-boson system this size is $N_\mathrm{conf}=\binom{N+M-1}{N}$,
where $M$ is the number of one-particle functions (orbitals) used to build the symmetrized Hartree products.
The two-boson problem can be diagonalized by taking a lot of basis functions,
while already for a ten-boson problem with $M=16$, $N_\mathrm{conf}=3,268,760$, meaning that the diagonalization of the respective 
secular matrix is not a simple task.
Due to this binomial dependency of the size of the spanned Fock subspace,
the bosonic system with large number of particles can be tackled only with quite a few orbitals, for example
for $N=1000$ and $M=3$ the size of the secular problem is $N_\mathrm{conf}=501,501$,
while already for $M=4$ it is $N_\mathrm{conf}=167,668,501$.
One of the main goals of the present work is to verify that the choice of the basis functions
(orbitals) used to build up the permanents (symmetrized Hartree products)
has enormous impact on the numerical convergence of many-body problems.

The above considered analysis of the interplay between 
the exact solution written in the center of mass and in the laboratory frames of references
is of course applicable for an arbitrary number of particles.
The Hartree products appearing in the solutions of the corresponding HIM problems written in the laboratory frame 
are build up of HO basis functions with exponents $-\frac{\omega+\delta_N}{4}x^2_j$, i.e., they
depend via $\delta_N$ on the number of particles $N$ and on the interparticle interaction strength $K_0$.
Hence, to solve the HIM problem for different parameters $N$ and $K_0$ in the laboratory frame
it is advantageous to use one-particle basis functions with different shapes (exponents).
It is natural to ask the following questions: 
What to do in a general case, when the analytic solution of the problem is not available, i.e., which basis set to use?
And how to find the best ``optimal'' self-consistent orbitals?
The simplest answer is to use the one-particle functions (bare orbitals) of the studied system
when the two-body interparticle interactions are switched off.
These fixed-shape basis functions are obtained as solutions of the one-particle problem 
$\hat h \psi_i=e_i \psi_i$ with $\hat h=\hat {T}(\vec{r}) + \hat{V}(\vec{r})$.
In the studied HIM model it means to use the HO basis $\psi^{ho}_n(x,\Omega)$ with trapping frequency $\Omega=\omega$.
The answer to the second question is also known -- to use the recently developed MCTDHB method \cite{streltsov:07,alon:08},
which utilizes the Dirac-Frenkel variational principle to determine the optimal shapes of the orbitals for
time-dependent problems.
In this work we will examine and contrast the performance of two many-body methods to attack the 
time-dependent and time-independent HIM problems 
-- the full configuration interaction (exact diagonalization) method which utilizes as a basis set the solutions
of the one-particle in the harmonic trap, and the self-consistent MCTDHB method.

\section{The MCTDHB Method}\label{MCTDHB_th}

Let us briefly describe the MCTDHB theory, for complete derivations and some recent applications see
the literature \cite{streltsov:07,alon:08,Streltsov2010,fragmentons,sakmann:09,sakmann.pra:10,
streltsov:PRL11,streltsov:PRA11,grond:09,grond:11a,grond:11b,brezinova:12}.
The MCTDHB method has been developed to solve the time-(in)dependent many-boson Schr\"odinger equation.
It relies on a multiconfigurational ansatz for the wavefunction, 
i.e., $\vert \Psi (t) \rangle=\sum_{\vec{n}} C_{\vec{n}}(t) \vert \vec{n}; t \rangle$.
The unknown $C_{\vec{n}}(t)$ are called expansion coefficients.
The permanents $\vert \vec{n}; t \rangle$ are build as symmetrized Hartree products of $N$
unknown orthogonal one-particle functions.
For $M$ orbitals the number of these permanents is equal to the number all possible permutations
of $N$ particles over $M$ orbitals, namely $\binom{N+M-1}{N}$. 
It is noteworthy that \textit{both} the coefficients $C_{\vec{n}}(t)$ and the one-partice functions used to build the permanents 
are time-dependent, variationally optimized quantities, which are determined 
by solving the corresponding MCTDHB equations of motion \cite{streltsov:07,alon:08}.
These equations depend on the parameters of the Hamiltonian, on the number of particles as well as on the number
of the one-particle basis functions used.
For different evolution times the optimal orbitals have different shapes -- this feature is called time-adaptivity.

Within the MCTDHB method the time-independent variational solutions are obtained by 
propagating the MCTDHB equations in imaginary time, which is
equivalent to solving the stationary problem variationally, 
as developed in the multiconfigurational Hartree method (MCHB) for bosons, see Ref.~\cite{MCHB}.
Hence, the static solutions we give here qualify as test suits for how the standard (time-independent) variational 
principle is handled numerically by MCTDHB.
From now on we call the time-dependent variational MCTDHB solution ``time-adaptive''
to distinguish it from the ``self-consistent'' static, i.e., time-independent MCTDHB solution.
If the one-particle functions used are not allowed for optimization, the MCTDHB method boils down to the
standard full configuration interaction (exact diagonalization) method.
Thus, one can consider the MCTDHB method as an exact diagonalization method with time-adaptive (self-consistent)
orbitals. For a given number of orbitals the dimension of the secular problem involved for the FCI(ED) and MCTDHB computations
is the same, $N_\mathrm{conf}=\binom{N+M-1}{N}$.
If only one self-consistent orbital is considered the MCTDHB theory boils down to the famous Gross-Pitaevskii (GP) mean-field theory
widely and often successfully used to describe static and dynamics of condensed bosonic systems
\cite{Pitaevskii_review,Leggett_review,pethick:08}.

At this point it is very important to stress that the MCTDHB and standard FCI methods used in this work
operate with general Hamiltonians in the laboratory frame of reference, so,
the separability of the HIM model is not taken into account.
But this formulation of the methods allows one to attack general, 
i.e., inseparable problems as has been done in, e.g.,
Refs.~\cite{sakmann:09,sakmann.pra:10,streltsov:PRA11,grond:09,grond:11a,grond:11b,brezinova:12}.

The MCTDHB equations of motion are solved numerically efficiently with the MCTDHB program package \cite{Streltsov2010}.
The current study relies on the propagation of the orbitals' equations of motion 
with a shared-memory parallelized implementation
of the Adams-Bashforth Moulton predictor corrector integrator \cite{recipesF:92}
and the coefficients' equations of motion with a hybridly OpenMP-MPI
parallelized short iterative Lanczos algorithm \cite{SIL}.
As primitive basis functions representing the self-consistent (time-dependent) orbitals 
we use either the HO discrete variable representation \cite{DVR} or 
the fast Fourier transform collocation method utilizing hybrid OpenMP-MPI parallelization, see \cite{Streltsov2010}.

\section{Ground state of the HIM: MCTDHB and FCI vs. exact solution}\label{GS}

\begin{figure}
\includegraphics[angle=-90,width=\textwidth]{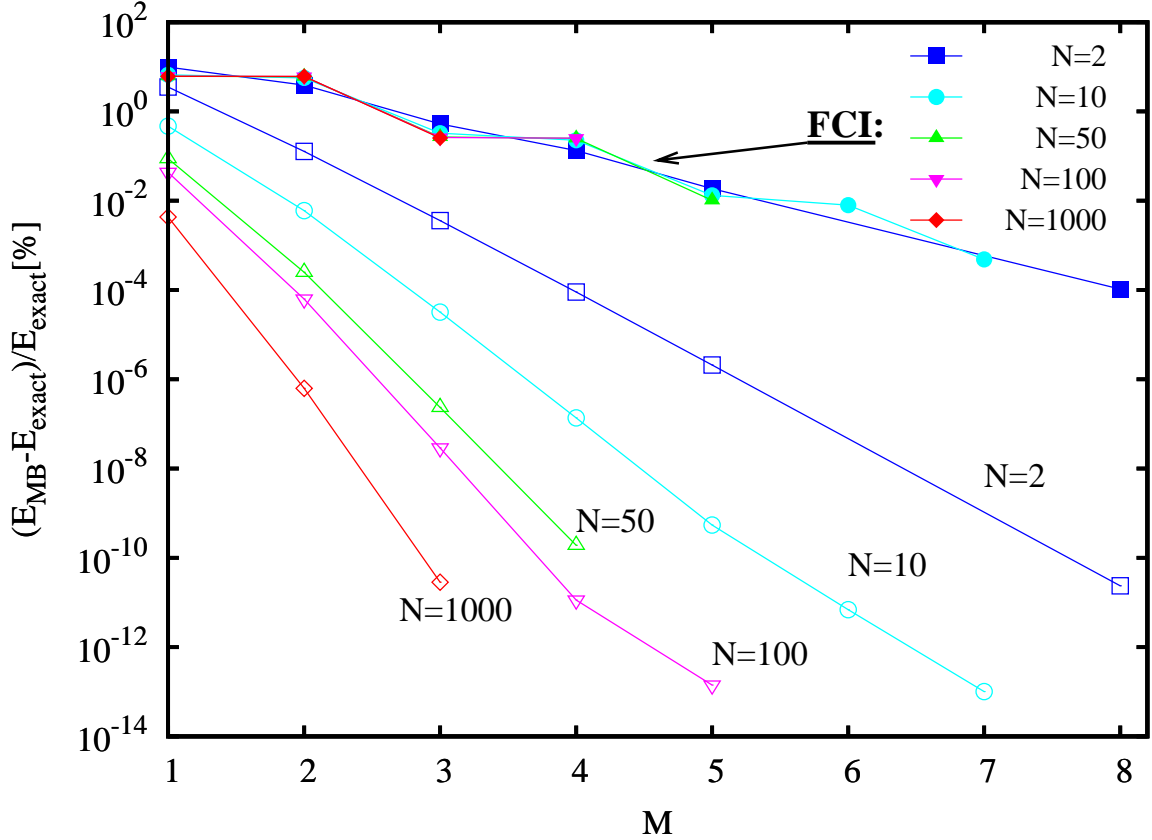}
\caption{
(color online).
Numerical convergence of the self-consistent MCTDHB and fixed-orbital full configuration interaction (FCI) methods
for the ground state energy of the harmonic interaction model (HIM).
Systems with N=2,10,50,100 and 1000 bosons are considered, the strengths of the interparticle interactions $K_0$
have been chosen to keep $\Lambda=K_0(N-1)=0.5$ constant. We plot the relative
differences between the total energies computed using the MCTDHB (filled symbols) and FCI (open symbols) many-body methods 
and respective exact energies in percents, $100 \cdot (E_{\mathrm{MB}} - E_{\mathrm{exact}})/E_{\mathrm{exact}}$, for different orbital number $M$.
For a given $M$ both many-body methods span the same Fock space, i.e., the respective 
secular matrices to be diagonalized are of the same size.
The advantage of the appropriate, i.e., self-consistent, choice of the one-particle basis functions is evident --
the self-consistent MCTDHB method converges much faster than the fixed-orbital FCI one.
Note the logarithmic scale and number of decades spaned.
All quantaties shown are dimensionless.}
\label{HIM_MCTDHB}
\end{figure}

We begin by benchmarking the MCTDHB and FCI methods against the ground state of the one-dimensional HIM.
We consider systems of $N=2,10,50,100,1000$ bosons trapped in the parabolic trap potential $V(x)=\frac{1}{2}x^2$
with the inter-boson interaction strengths selected to keep  $\Lambda=K_0(N-1)=0.5$ constant.
Such a choice of the interaction strengths implies that all these systems have the same GP solution,
i.e., are equivalent at the mean-field level of description.
To find the properties of convergence of the MCTDHB and FCI methods towards the exact solution of the HIM,
it is instructive to successively increase the number of orbitals, $M$, used in the computation.
In Fig.~\ref{HIM_MCTDHB} we plot the relative difference between the ground state eigenenergy and the corresponding exact energy $(E_{\mathrm{MB}}-E_{\mathrm{exact}})/E_{\mathrm{exact}}$ as a function of number of orbitals $M$ used.
The self-consistent many-body (MB) MCTDHB($M$) results are plotted by open symbols, the corresponding fixed-orbital
FCI$(M)$ results are depicted by filled symbols.

The key observation seen in Fig.~\ref{HIM_MCTDHB} is that the numerical results
converge towards the exact ones with increasing number of the orbitals used.
The performance of the self-consistent MCTDHB method, however, by far exceeds that of the fixed-orbital FCI.
Note the logarithmic scale and number of decades spanned!
The proper choice of the many-body basis set is very crucial -- within the same size of 
the Fock subspace spanned (dimension of the secular matrix $N_\mathrm{conf}$)
one can get an improvement of about six to eight orders of magnitude!
The results prove that the exact solutions of the HIM can be obtained numerically using the MCTDHB method
with just a few self-consistent orbitals.

Another striking feature of the MCTDHB method is its performance for different particle numbers.
The convergence is faster for larger particle numbers at fixed $\Lambda=K_0(N-1)$.
This is anticipated, because at the mean-field GP$\equiv$MCTDHB(1) level
the considered systems of bosons are equivalent and the GP solution of the static HIM problem
tends to an exact one in the thermodynamic ($N\to \infty$) limit.
Nevertheless, for large, but finite number of bosons the MCTDHB significantly improves the GP description.
For example, for N=1000 the relative differences to the exact energy obtained within 
the GP and MCTDHB(3) are $\sim10^{-3}$ and $\sim10^{-11}$ percents respectively, see Fig.~\ref{HIM_MCTDHB}.
So, the self-consistency becomes more and more relevant for systems made of larger particle numbers.
In contrast to the MCTDHB, the fixed-orbital FCI method does not show such a tendency.
The low performance of the full configuration interaction method utilizing the bare 
HO orbital basis set is evident from the above done two-boson analysis in Eq.~(\ref{HIMGS2bosons}).
Instead of the HO eigenfunctions of trap potential $\psi^{ho}_n(x,\Omega=\omega)$ 
one has to use the HO basis functions with ``modified'' frequency $\Omega=\frac{\omega+\delta_N}{2}$.
However, in the general case, when an analytical solution is unavailable
the only strict way to find the ``proper'' basis set is to solve the MCTDHB($M$) equations,
which determine variationally the optimal one-particle functions, see Ref.~\cite{MCHB}.

To highlight the convergence of the MCTDHB($M$) method with the number of orbitals $M$ used
we present in table \ref{tabHIMenergies} the total ground state energies of the above considered
systems of N=10,100,1000 bosons with $\Lambda=K_0(N-1)=0.5$.
The exact ground state energies are from Refs.~\cite{Cohen:85,Yan:03}, also see Eq.~(\ref{himEexact}).

\section{quenching the interparticle interaction: MCTDHB and FCI vs. exact results}\label{HIMquench}

In the previous section we have seen that the numerically exact ground state solutions of the HIM
can be obtained using the MCTDHB method with just a few self-consistent orbitals.
The standard full configuration interaction method utilizing the ``non-optimal'' fixed-shape orbitals
of the non-interacting system has demonstrated a much worser convergence.
It makes the usage of the direct diagonalization method with ``non-optimal'' orbitals
for large particle numbers impractical.
Within the number of orbitals technically allowed to be used (this number defines
the size of the respective secular matrix to be diagonalized)
the quality of the obtained many-body results is unsatisfactory.
They can be worser than the one-orbital self-consistent mean-field (GP) results, see Fig.~\ref{HIM_MCTDHB}.

Having established the great relevance of self-consistency for statics,
in the present section we clarify its impact on the quantum many-boson dynamics.
The main difference between statics and dynamics is that quantum dynamics involves a lot of excited states
and, therefore, the applied many-body method has to be capable to describe them.
Indeed, the evolution of any given initial many-body state is obtained as a solution
of the time-dependent Schr\"odinger equation:
\begin{eqnarray}
\Psi(t)&=&\sum_{j=0}^{\infty} a_{j} e^{-i E_j t} \Phi_j(\vec{X})  \nonumber \\
        &=&\sum_{j=0}^{\infty}  \langle \Psi_0(\vec{X})| \Phi_j(\vec{X}) \rangle e^{-i E_j t} \Phi_j(\vec{X}).
\label{TDEXPAND}
\end{eqnarray}
Here $\Psi_0(\vec{X})$ is the initial state and $\Phi_j(\vec{X})$ and $E_j$ are the eigenstates 
and respective eigenenergies of the quantum system, $\vec{X}$ are the coordinates of the constituting particles.
For the HIM considered here all the eigenstates and respective eigenenergies are known in the center of  mass frame,
so, to study the exact evolution of the many-body system one needs to evaluate 
the overlap integrals $a_{j}=\langle \Psi_0(\vec{X})| \Phi_j(\vec{X}) \rangle$.
When computing with the MCTDHB we, of course, work in the laboratory frame and the time-dependent many-body wavefunction is 
a complicated non-terminating expansion in terms of permanents.

Let us study a scenario with the HIM Hamiltonian where 
the many-body dynamics is activated by a sudden quench of the interparticle interaction strength.
It is worthwhile to mention that the MCTDHB method has been successfully used in Ref.~\cite{sakmann:10}
to describe such a scenario for ultra-cold systems with contact interaction.
On the experimental side the quench of the interparticle interaction
is a routine procedure controlled by the Feshbach resonance technique.
We assume that the initial state just before the quench
was the ground state of the non-interacting system.
What kind of dynamics is anticipated in this case?
The initial state, i.e., the ground state of the harmonically trapped system is symmetric,
implying that the one-body density has ``gerade'' symmetry. 
The sudden quench of the interparticle interaction cannot break this symmetry.
So, we expect that a change (quench) of the interparticle interaction leads to a ``breathing'' dynamics of the system --
the many-body wavefunction changes its shape such that the ``gerade'' symmetry is preserved.
This dynamical behavior is general and persists
in other many-body systems with symmetric trap potentials as well, e.g., in ultra-cold systems with contact interactions.

\subsection{Breathing dynamics for $N=2$ bosons}

We first consider the two-boson HIM system where the interparticle interaction strength $K_0$
is suddenly quenched from zero to $K_0=0.5$. The exact general eigenstates of the HIM system are
products of two HO wavefunctions -- one describes the motion of the relative $q_1=\frac{x_2-x_1}{\sqrt{2}}$ coordinate,
another the motion of the center of mass $q_2=\frac{x_2+x_1}{\sqrt{2}}$.
Clearly, the center of mass part always has bosonic symmetry -- it does not change sign when coordinates of 
the particles are permuted. In contrast, the relative part can be either bosonic (symmetric)  or fermionic (antisymmetric),
depending on the parity of the Hermite polynomial involved.
For example, the first excited HO state of the relative part 
$\psi^{ho}_1(q_1,\tilde \omega)\sim H_1(\sqrt{\tilde \omega} q_1)e^{-\tilde \omega \frac{q_1^2}{2}} \equiv \mathcal{N} q_1 e^{-\tilde \omega \frac{q_1^2}{2}}$
[$\tilde \omega=\frac{\omega+\delta_2}{2}$] is fermionic, because the permutation $x_1 \leftrightarrow x_2$ changes 
the sign of the $q_1=\frac{x_2-x_1}{\sqrt{2}}$ (first Hermite polynomial) and,
therefore, the overall sign of the total wavefunction $\psi^{ho}_1(q_1,\tilde \omega)\psi^{ho}_0(q_2,\omega)$.
Using this argumentation one can conclude that
all even excited HO states $\psi^{ho}_i(q_1,\tilde \omega)$ with $i=0,2,4,...$
of the relative part are bosonic and all odd ones, i.e., $\psi^{ho}_i(q_1,\tilde \omega)$ $i=1,3,5,...$ are fermionic.

Having understood the nature of the bosonic and fermionic solutions of the two-particle HIM problem 
we are ready to analyze the excitations responsible for the ``breathing'' dynamics.
The ``ungerade bosonic excitations'' of the HIM model are activated by odd excitations of the center of mass part
$\psi^{ho}_i(q_2,\omega)$, which oscillates with the original trap frequency $\omega$.
The lowest ``gerade'' excitation corresponds to the second excited state of the relative part,
the next ``gerade'' bosonic excitation -- to the fourth excited state and so on.
In principle, the next class of the ``gerade'' excitations appears by product of the HO solutions 
corresponding to the second excited state of the center of mass motion
and every even excitation of the relative motion.
In the studied quench dynamics we start from the ground state of the non-interacting system,
implying that the center of mass motion is in the ground state.
Therefore, all excited states for which the center of mass is excited are orthogonal
to such an initial state and, hence,  do not contribute to the dynamics;
this is because the overlap integrals of these states with the initial state are zero.
Summarizing, the sudden quench of the interparticle interaction in the HIM leads to the breathing dynamics
with breathing frequencies 
$\omega(n)=2n\sqrt{\omega^2+4K_0}$, with main excitation frequency $\omega(n=1)\equiv\omega_{breath}$
and all its overtones with $n=2,3,4,...$. These frequencies are obtained as energy differences between 
the ground and respective excited eigenstates.

\begin{figure}
\includegraphics[angle=-90,width=8cm]{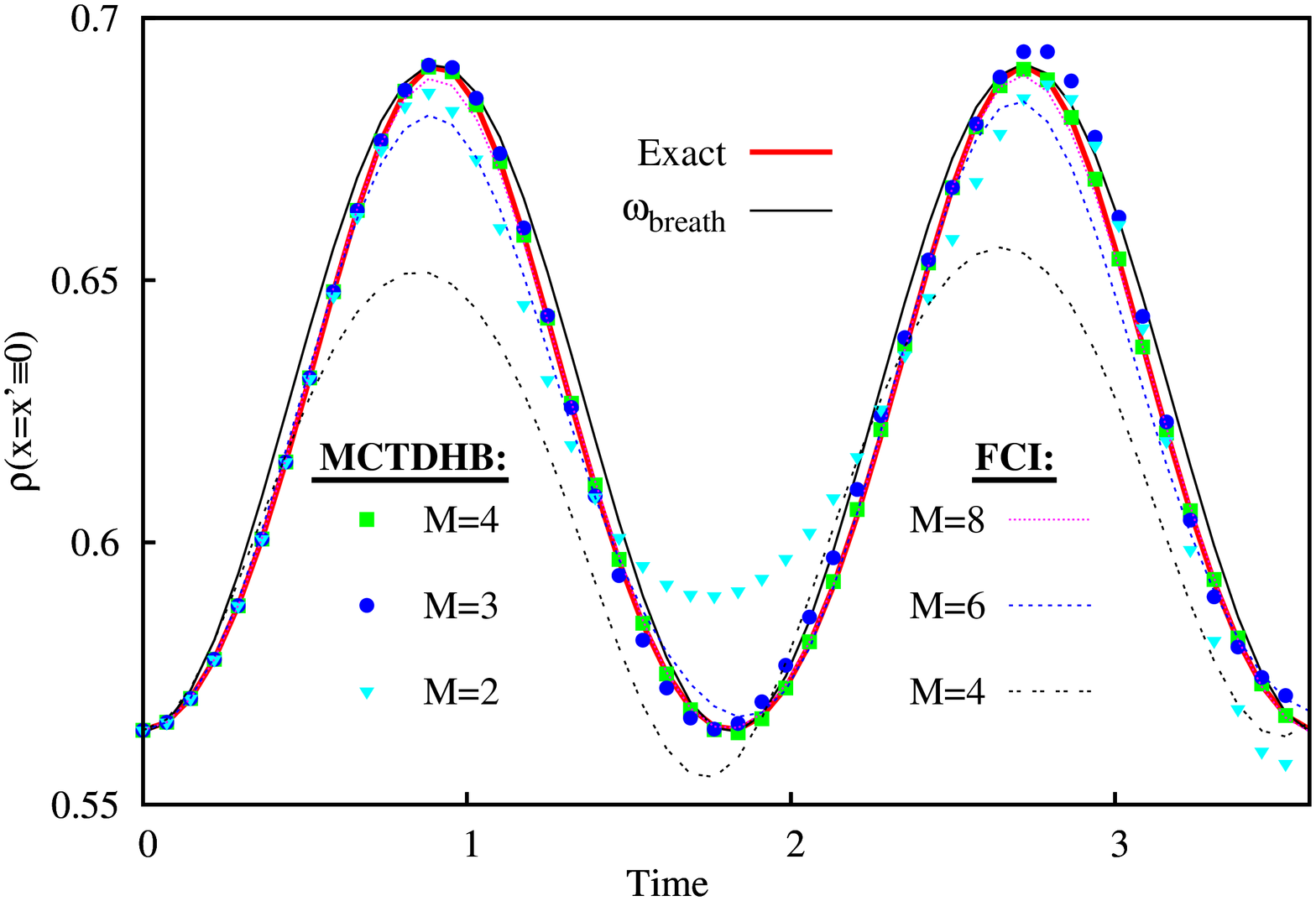}
\includegraphics[angle=-90,width=8cm]{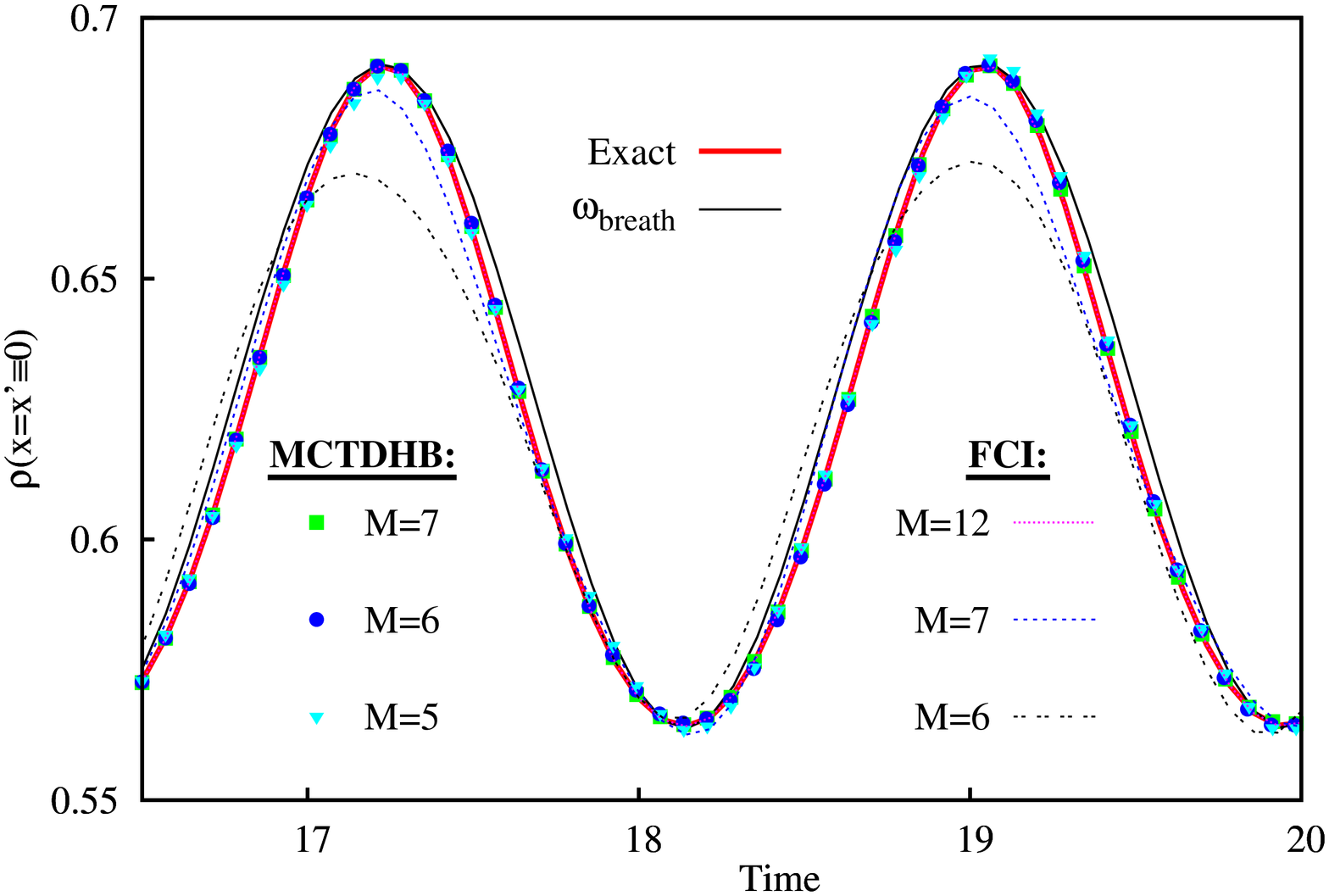}
\caption{
(color online).
A sudden change (quench) of the interparticle interaction leads to ``breathing'' dynamics of the system.
We study the HIM system with $N=2$ bosons where the interparticle interaction strength is quenched from zero to $K_0=0.5$.
The evolution of the one-particle density at the origin $\rho(x=x'\equiv0)$ is plotted as a function of time.
The exact dynamics reveals oscillations with the main breathing frequency $\omega_{breath}=2 \sqrt{\omega^2+4 K_0}$
augmented by overtones $2\omega_{breath}$, $3\omega_{breath},\ldots$ (see text for more details).
The solid (black) line depicts the guiding $\sim\cos(w_{breath}t)$ function.
The numerical MCTDHB and FCI results are contrasted and compared with the exact ones, plotted by a bold (red) line.
The left panel depicts the density oscillations at short times.
At the FCI level accurate description of the dynamics is achieved by using at least eight fixed-shape orbitals.
To gain similar accuracy within the MCTDHB method one needs only three time-adaptive orbitals.
Four time-adaptive orbitals [MCTDHB($M=4$)] provide a numerically exact description.
The right panel shows the density oscillations at longer times.
To describe the dynamics in this case a larger Fock space (more orbitals) is required.
The numerically exact description is obtained by using six time-adaptive [MCTDHB(6)] or twelve fixed-shape orbitals [FCI(12)].
See text for further discussion. All quantities shown are dimensionless.} 
\label{HIM_QUENCH_N2}
\end{figure}

We use the Mathematica package \cite{Mathematica} to compute the required overlap integrals
in Eq.~(\ref{TDEXPAND}) and to get the exact time-dependent two-boson wavefunction.
Here we have to mention that instead of the infinite summation
the contributions from the 60 exact lowest-in-energy excited states are taken into account; this is more than sufficient
for numerical convergence.
Next, the exact one-body density as a function of time is obtained according to its definition:
the two-body wavefunction is multiplied by its complex conjugate and one coordinate is integrated out.
In Fig.~\ref{HIM_QUENCH_N2} we plot the exact value of the density at the trap center as a function of time by a bold red line.
The first two breathing cycles are depicted in the left panel of this figure,
the right panel presents the breathing dynamics at longer propagation times.
The numerical $M$-orbital MCTDHB and FCI results are depicted by bold symbols and dotted lines, respectively.

The first observation seen from Fig.~\ref{HIM_QUENCH_N2} is that the exact density at the middle of the trap oscillates
periodically with the breathing frequency $\omega_{breath}=2\sqrt{\omega^2+4K_0}$.
However, the shape of the oscillation differs from the simple $\sim\cos{(\omega_{breath}t)}$ function plotted to guide the eye
by a solid black line. This is the result of the contributions from the overtones originating from the higher excited states.
The two-orbital MCTDHB(2) solution provides essentially an exact description of the dynamics till 
half of the breathing cycle -- notice the triangles following the exact results.
The three-orbital MCTDHB(3) results, plotted by filled circles, are on-top of the exact curve for the first breathing cycle;
small deviations from the exact results become visible at the second breathing cycle.
The MCTDHB(4) with four time-adaptive orbitals gives the exact description of the first two breathing cycles.
The FCI dynamics with four fixed-shape orbitals, plotted by a double-dashed line,
starts to deviate from the exact results already at very short times.
Even the six-orbital FCI dynamics, depicted by a dashed line, starts to deviate from the exact result
after one third of the first breathing cycle.
A quite accurate description of the first two breathing oscillations is only obtained on the eight-orbital FCI level.
Summarizing, to describe the first two breathing oscillations of the two-particle HIM problem
one needs either four time-adaptive orbitals [MCTDHB(4)] or eight fixed-shape orbitals [FCI(8)].
The above analysis also shows that with time more time-adaptive orbitals are needed to describe
the exact dynamics.

Now we quantify the performance of the MCTDHB and FCI methods to describe the quantum dynamics at longer times.
In the right part of Fig.~\ref{HIM_QUENCH_N2} we plot the oscillations of the density at the middle of the trap
at longer times. Exact results are depicted by a bold line, the MCTDHB results are depicted by filled symbols
and the FCI results are plotted by dashed lines. The exact density continues to oscillate with the main $\omega_{breath}$
and its overtones. The MCTDHB($M>5$) results are numerically exact. The FCI(12) result,
plotted by a dense-dashed line, follows the exact one, while the FCI(6) and FCI(7) are clearly off.

Concluding, for longer propagation time one has to span a larger Fock subspace, i.e., one has to use a larger number 
of one-particle functions to build the permanents.
The difference between the quantum dynamics utilizing fixed-shape and time-adaptive orbital basis sets used is clearly seen --
to gain a desired accuracy of the propagation one needs to use at least twice as many fixed-shape orbitals than time-adaptive ones.
The second important observation is that if a desired convergence of the many-body dynamics is achieved 
on the $M$-orbital level, further extension of the Fock space is unnecessary, the inclusion of the extra orbitals 
does not impact the result. This is a general consequence of the variational principle used.
It is known for the FCI method and now proven for the MCTDHB method, which is based
on the time-dependent Dirac-Frenkel variational principle.
This feature allows us to define a practical strategy for MCTDHB computations.
If the dynamics obtained on the MCTDHB($M$) and MCTDHB($M+1$) levels are identical we conclude that
numerical convergence to the exact results is reached.
In other words, the many-body wavefunction built from $M$ time-adaptive orbitals is the converged solution
of the time-dependent many-boson Schr\"odinger equation.

\subsection{Breathing dynamics for $N=10$ bosons}

\begin{figure}
\includegraphics[angle=-90,width=\textwidth]{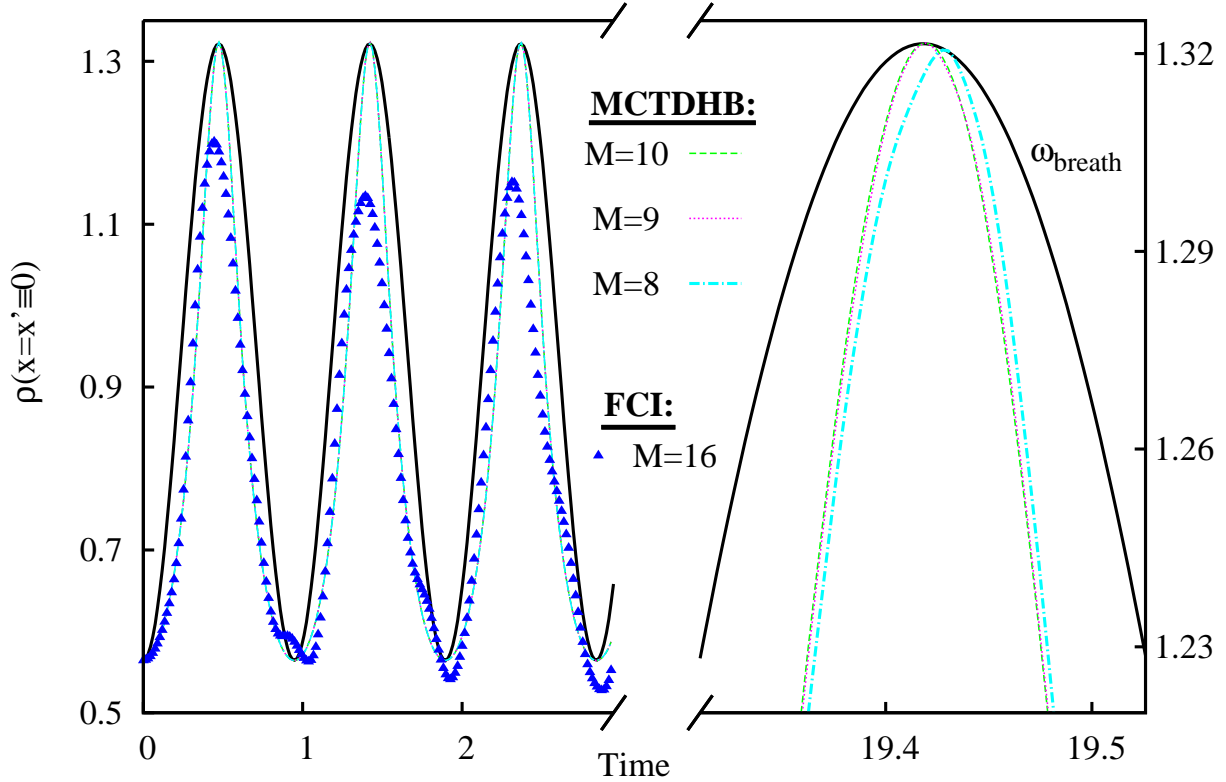}
\caption{(color online).
Breathing dynamics of the HIM system with $N=10$ for the same interaction quench scenario as in Fig.~\ref{HIM_QUENCH_N2}.
The evolution of the one-particle density at the origin $\rho(x=x'\equiv0)$ is plotted,
notice different scales for the short and long times.
The density oscillation is formed by the main breathing frequency $\omega_{breath}=2 \sqrt{\omega^2+2 N K_0}$
with strong contributions of the overtones $2\omega_{breath},3\omega_{breath},...$ (see text for more details).
The solid (black) line depicts the guiding $\sim\cos(w_{breath}t)$ function.
The MCTDHB($M=8$) method with eight time-adaptive orbitals provides very accurate description of the breathing dynamics for
the short and long times. The MCTDHB results for $M=9$ and $M=10$ are identical, indicating that the exact description
has been numerically reached.
The FCI($M=16$) results plotted by triangles start to deviate from the exact solution already for short times.
The FCI method with sixteen fixed-shape orbitals provides a reasonable description 
of the dynamics for a very short time only, i.e., it is incapable to describe more than a half of the first breathing cycle.
The exact results could not be obtained in this model from the analytical solution Eq.~(\ref{TDEXPAND}) because
it is more difficult to perform the needed 10-dimensional integrations than to solve the problem numerically
exactly by MCTDHB.
All quantities shown are dimensionless.} 
\label{HIM_QUENCH_N10}
\end{figure}

Now we examine and compare the performance of the MCTDHB and FCI methods to treat the time-dependent dynamics of
the HIM system with N=10 bosons for the same quench scenario as studied before for a system with N=2 bosons.
By analyzing the structure of the excited states for the $N=10$ system we arrive at the conclusion that a sudden quench
leads to the many-body breathing dynamics with frequencies $\omega(n)=2n\sqrt{\omega^2+2NK_0}$,
obtained as energy differences between the ground and respective excited eigenstates.
Here, the lowest excitation $\omega(n=1)\equiv\omega_{breath}$ is responsible for the main breathing excitation frequency,
higher excited states with $n=2,3,4,...$ result in overtones.
The exact results are in principle available for the systems with any number of particles \cite{Cohen:85,Yan:03}.
However the straightforward way of evaluation of the exact time-dependent many-body wavefunction and the respective density 
successfully applied for two-boson is much more involved for the ten-particle system.
Hence, we employ the numerical MCTDHB method.
The first three breathing cycles are depicted in the left part of Fig.~\ref{HIM_QUENCH_N10}.
The MCTDHB results utilizing $M=8,9,10$ time-adaptive orbitals plotted by dashed lines are indistinguishable from each other.
The computational strategy verified and proven above allows us
to conclude that the numerically exact description of the TDSE is achieved.

Another important observation seen in Fig.~\ref{HIM_QUENCH_N10}  is that the numerically exact MCTDHB results 
deviate substantially from a simply fitted $\sim\cos{(\omega_{breath}t)}$ curve, plotted by a solid black line to guide the eye.
This is direct evidence that the contribution from higher excited states, responsible for higher overtones,
to the breathing dynamics of the $N=10$ boson system is much stronger than it was in the $N=2$ system
studied before [compare the exact curve and fit to the $\sim\cos(\omega_{breath}t)$ curve in Fig.~\ref{HIM_QUENCH_N2}].
The FCI result with $M=16$ fixed-shape orbitals depicted by filled triangles follows the numerically exact MCTDHB curves
only for a very short initial time -- for about one half of the first breathing cycle.
For longer propagation times the FCI(16) predictions deviate from the exact results.

The right part of Fig.~\ref{HIM_QUENCH_N10} depicts on an enlarged scale the breathing dynamics at longer times.
At all plotted propagation times the eight-orbitals MCTDHB(8) method provides a very accurate description
of the many-boson dynamics while the MCTDHB computations with $M=9,10$ time-adaptive orbitals are numerically exact.
We conclude that the usage of time-adaptive orbitals provides an enormous benefit for the accurate description
of quantum dynamics of systems with larger particle numbers.
In contrast to the MCTDHB method which is capable to provide numerically exact results with a few time-adaptive orbitals,
the FCI treatments even with much larger Fock subspaces spanned can not provide even a qualitative description of the dynamics.

\section{The time-dependent HIM: Non-equilibrium dynamics}\label{TDHIM}

The above discussed visualization of the HIM system
as a medium made of $N-1$ non-interacting ``relative'' particles
in which the effective mass particle (representing the center of mass coordinate)
lives, allows for a simple physical time-dependent generalization.
Without loss of the separability one can assume that
the effective-mass particle moves now not in the stationary 
but in a time-dependent harmonic potential with driving frequency $\omega(t)$.
Moreover, we assume that during this motion the medium representing the relative coordinates remains undisturbed $\delta_N=const.$.
Surprisingly, the Hamiltonian corresponding to this problem takes on a simple form in the center of mass frame:
\begin{equation}
\hat{H}(t)=\hat{H}_{rel}+\hat{H}_{CM}(t)=
\sum_{i=1}^{N-1}(-\frac12\partial^2_{\vec{q}_i}+\frac12\delta^2_N \vec{q}^2_i)-\frac12\partial^2_{\vec{q}_N}+\frac12 \omega^2(t) \vec{q}^2_N.
\label{omtranstd1}
\end{equation}
One can apply a reverse engineering and transform this new time-dependent HIM problem
back to the laboratory frame:
\begin{equation}
\hat{H}(t)=\sum_{i=1}^N [- \frac{1}{2}\partial_{\vec{r_i}}^2 + \frac12 \omega(t)^2 \vec{r_i}^2 ] +  K(t) \sum_{i<j}^{N} \left( \vec{r}_i - \vec{r}_j \right)^2.
\label{omtranstd2}
\end{equation}
This coupled time-dependent Hamiltonian corresponds to the situation where all ``real'' particles
are trapped in the time-dependent potential $\hat{V}(\vec{r},t)=\frac12 \omega(t)^2 \vec{r}^2$ and
interact via time-dependent harmonic interparticle interaction potential of strength $K(t)$
[which depends on $\omega(t)$ and $\delta_N$].
For the external trapping potential driven by a time-dependent function $f(t)$:
\begin{equation}
\omega(t)=\omega_0\left[1+f(t)\right].
\label{omtrans}
\end{equation}
the imposed above requirement $\delta_N=\sqrt{\omega_0^2 + 2 N K_0}=const.$
implies that the interparticle interaction strength has to be driven with the ``compensating'' time-dependency:
\begin{equation}
K(t) = K_0 \left[1 - \frac{\omega_0^2}{2NK_0} f(t)\right].
\label{Ktrans}
\end{equation}
Since the Hamiltonian (\ref{omtranstd1}) or (\ref{omtranstd2}) is now time-dependent,
the total energy is, of course, no longer conserved.

Let us consider a situation where the medium representing $N-1$ relative particles is in the ground state
of the harmonic potential with frequency $\delta_N = \sqrt{\omega_0^2 + 2 N K_0}$.
Its energy is the time-independent constant $\frac{D}{2}(N-1)\delta_N$.
The time-dependency of the full problem then originates from the driving of the center of mass:
\begin{equation}
\hat{H}_{CM}\psi(\vec{q_N},t)= -\frac12\partial^2_{\vec{q_N}}\psi(\vec{q_N},t) + \frac12 \omega^2_0[1+f(t)]^2 \vec{q_N}^2\psi(\vec{q_N},t)
= i \frac{\partial}{\partial t}\psi(\vec{q_N},t).
\label{TDSE1p}
\end{equation}
The solution $\psi(\vec{q_N},t)$ of this {\it one-particle} Schr\"odinger equation can easily 
be obtained numerically, see Refs.~\cite{kosloff:88,recipesF:92}.
The final expression for the expectation value of the Hamiltonian Eq.~(\ref{omtranstd1}) or ~(\ref{omtranstd2}) reads:
\begin{equation}
 \langle \Psi(t) \vert \hat{H}_{rel}+\hat{H}_{CM} (t) \vert \Psi(t) \rangle= \frac{D}{2}(N-1)\delta_N+\epsilon(t),
\label{TDEXPVALE}
\end{equation}
where $\epsilon(t)=\langle\psi(\vec{q_N},t)|\hat{H}_{CM}(t)|\psi(\vec{q_N},t)\rangle$.
Interestingly, the special kind of time-dependency used in Eqs.~(\ref{omtrans}, \ref{Ktrans})
also implies that the time-dependent part $\epsilon(t)$ of
the expectation value of the total Hamiltonian $\hat{H}(t)$
depends neither on the number of particles $N$ nor on the interaction strength $K_0$.
So, the systems with different particle numbers $N$ and different interparticle interaction strengths $K_0$
possess the same time-dependent fraction.

It is instructive to state here that the time-dependencies, Eqs.~(\ref{omtrans},\ref{Ktrans}),
can be more general and it is of course not necessary to choose them such that
the relative Hamiltonian $\hat{H}_{rel}$ remains time-independent, i.e., keeping $\delta_N=const.$ 
Yet, there is one important advantage to this particular choice:
in the center of mass frame, Eq.~(\ref{omtranstd1}), the relative part is known analytically and to solve the problem completely and exactly
one needs to integrate only a single one-particle Schr\"odinger equation, Eq.~(\ref{TDSE1p}).
In contrast, to find the solution in the laboratory frame one has to solve the time-dependent many-boson Schr\"odinger equation
with a {\it time-dependent trap potential} and {\it time-dependent interparticle interactions}.
While the former task is a manageable standard routine, the latter one comprises a very involved and appealing
theoretical and numerical problem.
The main goal of the present section is to show that the MCTDHB method is capable to tackle time-dependent scenarios
numerically exactly even in the most involved setups: time-dependent one-particle potentials $\hat{V}(\vec{r},t)$ and time-dependent two-body interactions $\hat{W}(\vec{r},\vec{r}',t)$.

In what follows, we investigate the dynamics of the one-dimensional HIM system
with time-dependent trapping (\ref{omtrans}) and interaction (\ref{Ktrans}) potentials
driven by two different functions
\begin{eqnarray}
f_1(t)&=&0.2 \sin^2(t), \nonumber \\
f_2(t)&=&\sin(t)\cos(2t)\sin(0.5t)\sin(0.4t).
\label{f1_f2}
\end{eqnarray}
The one-body center-of-mass Schr\"odinger equation (\ref{TDSE1p}) with the respective time-dependent potentials
is integrated numerically exactly to obtain the corresponding one-body energies $\epsilon_1(t), \epsilon_2(t)$.

\begin{figure}
\includegraphics[angle=-90,width=\textwidth]{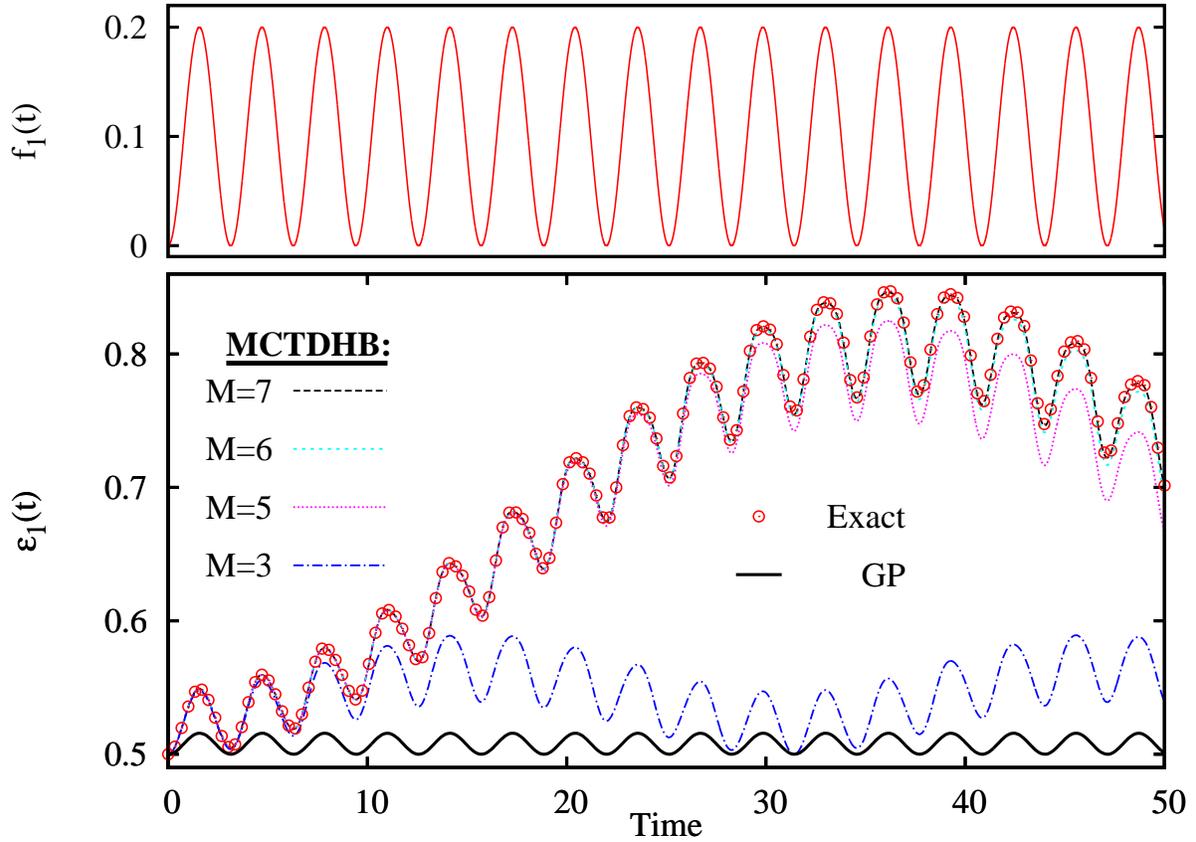}
\caption{
(color online).
The HIM model with time-dependent trap $V(x)=\omega_0\left[1+f(t)\right]x^2$
and time-dependent interparticle interaction $W(x_i-x_j)=K_0 \left[1 - \frac{\omega_0^2}{2NK_0} f(t)\right](x_i-x_j)^2$ permits
exact solution. The exact expectation value of the total Hamiltonian of the system Eq.~(\ref{TDEXPVALE}) reads
$\langle \Psi(t) \vert \hat{H}(t) \vert \Psi(t) \rangle= \epsilon(t) + const.$,
with the time-independent constant equals to $\frac{D}{2}(N-1)\delta_N$ and $D=1$.
The driven function $f(t)=f_1(t)$ and the time-dependent part of the energy
$\epsilon(t)=\epsilon_1(t)$ for N=10 bosons with $K_0=0.5$ are plotted.
The convergence of $\epsilon_1(t)$ when increasing the number of the time-adaptive orbitals $M$ is depicted.
The Gross-Pitaevskii results [GP $\equiv$MCTDHB(1)], plotted by a bold solid line, are inaccurate even for very short time.
The MCTDHB(3) provides excellent description up to $t\approx5$, the MCTDHB(5) works well till $t\approx 15$,
the MCTDHB(6) till $t\approx 30$, and the MCTDHB(7) results coincide with the exact solution at all the times depicted.
See text for discussion. All quantities shown are dimensionless.}
\label{HIMTD_ft1}
\end{figure}

Let us first study the time-dependent HIM system made of $N=10$ bosons with
a relatively simple periodic driving function $f_1(t)$ and $K_0=0.5$.
In the lower part of Fig.~\ref{HIMTD_ft1} we plot the time-dependent part $\epsilon_1(t)$ 
of the respective expectation value of the total Hamiltonian $\hat{H}(t)$ computed by using different levels of the MCTDHB($M$) theory.
The numerically exact results for $\epsilon_1(t)$ depicted by open circles
are obtained by solving directly the one-particle time-dependent Schr\"odinger equation.
It is important to notice that the oscillatory motion of the 
center of mass results in a relatively small contribution to the total 
energy: the value of $\epsilon_1(t)$ is of the order of a single-particle energy.

The close inspection of Fig.~\ref{HIMTD_ft1} shows that
the MCTDHB(3) computation, depicted by a dashed-doted line, follows the exact curve until $t\approx5$.
To describe the exact dynamics for longer propagation times one needs to use higher levels of the MCTDHB($M$) theory,
i.e., more time-adaptive orbitals are needed.
The MCTDHB(5) result, plotted by a dense-dashed line, is exact until $t\approx15$, while
the MCTDHB(6) one depicted by a simple dashed line is exact until $t\approx30$.
The double-dashed line depicting the MCTDHB(7) result reproduces the exact 
time-dependency of the total energy at all the times considered here.

In the context of ultra-cold physics the Gross-Pitaevskii mean-field theory is considered as one of the main working tool
to describe the dynamics of bosonic systems with time-dependent traps and time-dependent interactions.
In Fig.~\ref{HIMTD_ft1} we also plot the results obtained by solving GP equation with the finite-range harmonic interaction,
which is identical to the lowest level MCTDHB(1) theory, by a bold solid line.
The GP theory is incapable to describe the time-dependent energy correctly even for short times. Note that $N=10$ only.

\begin{figure}
\includegraphics[angle=-90,width=\textwidth]{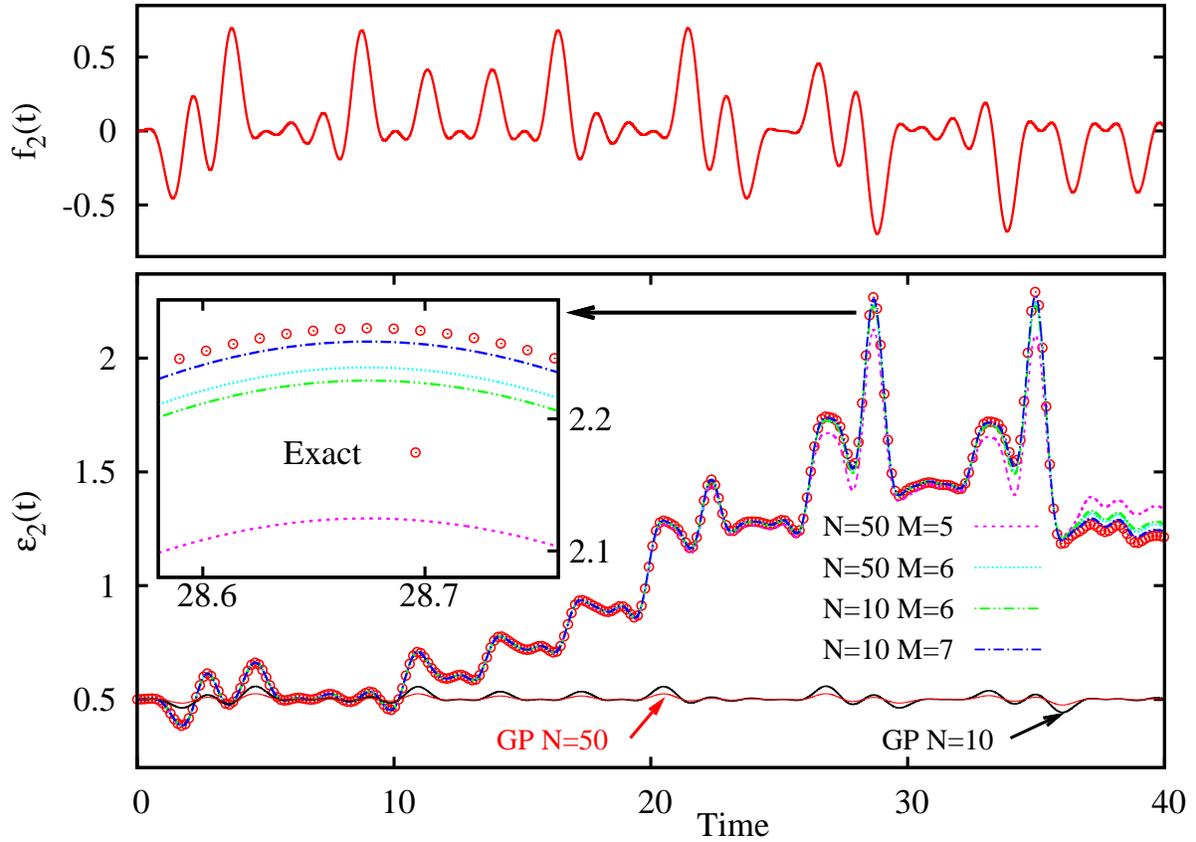}
\caption{(color online). The modified HIM model with time-dependent trap and interparticle interaction driven by a
complicated function $f_2(t)$ [Eq.~(\ref{f1_f2})]. The function is depicted in the upper panel.
The time-dependent contribution $\epsilon_2(t)$ to the total energies is computed
at several different levels of the MCTDHB($M$) theory for $N=10$ [$M=6,7$] and $N=50$ [$M=5,6$] bosons.
The strength of the interparticle interaction is $K_0=0.5$.
The considered time-dependency of the one- and two-body interaction potentials guarantees that
the exact $\epsilon_2(t)$, plotted by open red circles, is the same for both systems.
The MCTDHB(5) for $N=50$ and MCTDHB(6) for $N=10$ provide converged description of the dynamics till $t\approx25$,
for longer times more orbitals are needed for ``absolute'' convergence.
The corresponding Gross-Pitaevskii results, marked by arrows,
are semi-qualitative for very short initial times only, till $t\approx1$.
See text for discussion. All quantities shown are dimensionless.}
\label{HIMTD_ft2}
\end{figure}

Now we study the HIM systems made of N=10 and N=50 bosons with $K_0=0.5$
driven by quite a complicated time-dependent function $f_2(t)$, depicted in the upper part of Fig.~\ref{HIMTD_ft2}.
The exact $\epsilon_2(t)$, obtained by solving the corresponding one-particle Schr\"odinger equation,
is plotted by open red circles. As it was discussed above, this time-dependent fraction of the 
expectation value of the Hamiltonian $\hat{H}(t)$ is the same for both systems.
We use different levels of the MCTDHB($M$) theory to compute the time-dependent contribution
$\epsilon_2(t)$ to the energy of the studied systems.
The dashed-dotted and dashed-double-dotted lines are used, respectively, to depict the MCTDHB($M=6$) and
MCTDHB($M=7$) results for the system with $N=10$ bosons.
The MCTDHB($M=5$) and MCTDHB($M=6$) results for the system made of $N=50$ bosons
are depicted by the dashed and dense-dashed lines, correspondingly.
All the presented numerical MCTDHB results with $M>5$ follow the exact lines till $t\approx 25$,
indicating that numerical convergence is reached.
In Fig.~\ref{HIMTD_ft2} we also depict the corresponding Gross-Pitaevskii results.
The GP theory, usually considered to be applicable for systems made of a larger number of particles,
provides a semi-qualitative description of the very short initial dynamics ($t \approx 1$),
afterwards its predictions sharply deteriorates.

Summarizing, the MCTDHB($M$) computations with a {\it given} number of time-adaptive orbitals
start to deviate from the exact result with time, see Fig.~\ref{HIMTD_ft2} and its inset.
The time-dependent variational principle used in the MCTDHB method implies that
the MCTDHB computations done with a larger number of the time-adaptive orbitals remain ``on-top'' of the exact curve
for {\it longer} propagation times. Even when the numerical many-body results slightly deviate form the exact at longer
propagation times they are quantitative and quite accurate 
-- all the spectral features of the exact behavior are reproduced, see Fig.~\ref{HIMTD_ft2}.
In conclusion, the MCTDHB method is capable to provide numerically converged results 
for time-dependent Hamiltonians with very general driving scenarios,
where both the external trap and interparticle interactions are driven in quite a complicated way.

\section{Discussion and Outlook}\label{sum}

In the present work we have compared the quantum many-boson physics of the HIM system
described in the laboratory and in the center of mass frames.
In contrast to the center of mass frame where the HIM problem is exactly solvable,
in the laboratory frame one has to apply numerical efforts to reproduce the exact results,
even in the two particle case. The relevance of self-consistency and time-adaptivity is demonstrated.
To solve time-independent problems, the standard many-body 
full configuration interaction (exact diagonalization) method 
requires to use a large number of fixed-shape (non-optimal) one-particle basis functions, 
thereby restricting its applicability to few-particle systems.
The usage of the MCTDHB method utilizing variational self-consistent basis set
allows to attack systems with larger particle numbers.
To verify how good the obtained static solution is one can use the straightforward methodology:
by comparing the eigenstates obtained by imaginary time propagation of the MCTDHB equations with M and M+1 orbitals 
one can conclude that numerical convergence is achieved. See table \ref{tabHIMenergies} for reference.

To check the relevance of the time-adaptivity we have first studied the time-dependent HIM problem where the dynamics
are initiated by a sudden quench of the interparticle interaction strength from zero to some finite value.
It has been shown that the many-body method (MCTDHB) utilizing variationally optimal time-adaptive orbitals
allows to obtain the numerically exact solutions 
for much longer propagation times in comparison to the fixed-orbital FCI method 
spanning a Fock space of the same size.
The methodology in determining the accuracy of the time-dependent solution obtained 
is to compare the properties of the MCTDHB solutions computed by using M and M+1 time-adaptive orbitals
at different propagation times. If more time-adaptive orbitals are used than needed,
the Dirac-Frenkel variational principle
keeps the superfluous orbitals unoccupied, i.e., they do not
contribute to the now converged and exact many-boson wavefunction.
Generally, we have found that 
one needs less time-adaptive orbitals to converge the results for bosonic systems with increasing number $N$
of particles when the interaction strength $\Lambda=K_0(N-1)$ is kept fixed.

In the broader context several other methods of the family of the multiconfigurational methods have been benchmarked in the field.
The first of these, the multiconfigurational time-dependent Hartree (MCTDH, MCTDHB's mother method) \cite{meyer:90,manthe:92,meyer:09},
was benchmarked with standard wave-packet propagation \cite{meyer:90,manthe:92,meyer:09}
as well as with experimental spectra (see, e.g., Refs.~\cite{raab:99/2,MCTDHbench:09}). 
The MCTDH for fermions (MCTDHF, a sister method of MCTDHB) \cite{scrinzi:03,scrinzi:05,kato:04,nest:05} 
was benchmarked with direct numerical solutions of the Schr\"{o}dinger equation
(see, e.g., Refs.~\cite{MCTDHFbench:10,MCTDHFbench:11}).
Similarly, the aim of the present study was to benchmark and assess 
the properties of the convergence of MCTDHB with respect to the number of variational parameters used.
Throughout this work, the MCTDHB method has been benchmarked with the standard HIM.
The convergence of the ground state and non-equilibrium dynamics has been demonstrated.
We prove, thereby, that the MCTDHB can be used to obtain \textit{numerically exact} solutions of the many-boson TDSE.

We have also shown that the exactly-solvable many-body HIM problem can be extended
to a driven time-dependent Hamiltonian.
Namely, if the time-dependent modulation of the harmonic trap 
is accompanied by the modulation of the interparticle interaction with the same driving function
(with a different amplitude) all the internal excitations in such a system can be compensated and
the many-body system behaves as a driven single particle system.
For systems with large particle numbers this driving contribution is of the order of a single-particle energy.
Physically, it means that the modulation of the harmonic trap can be almost completely compensated
by the corresponding modulation of the interparticle interaction.

The driving scenario proposed for the HIM model is based on the separability of the relative and center of mass
coordinates and, therefore, it can also be adapted to other many-body system with such a separability.
In particular, it can work in many-body systems trapped in harmonic potentials 
and interacting via other two-body potentials which depend on the interparticle separation.
However it should be noted that, whereas the ``compensating'' relation between the trap 
and interparticle modulations are of simple form for harmonic interactions,
see Eqs.~(\ref{omtrans},\ref{Ktrans}), in the case of ultra-cold gases (with contact interactions)
this relation is expected to be much more involved.
Summarizing, in trapped many-particle systems where the center of mass is separable 
a novel phenomena of ``dynamical compensation'' can take place 
-- all the excitations originating from a driven trapping potential can be  
almost completely dynamically compensated by the respective driving of the interparticle interaction potential.

The driven many-body system can in principle be realized in the context of ultra-cold physics.
It would correspond to an experimental setup where the trap potential (magneto-optical trap) and external magnetic
field, used by the Feshbach resonance technique, are driven such that
the relative phase and amplitude of the time-dependent modulations can be tuned.
By measuring, e.g., the density response as a function of the amplitude of the modulation applied
one can scan for and verify the predicted effect.
If the dynamical compensation does not take place,
more and more excited states would contribute to the dynamics, and in time 
the density will oscillate with larger and larger amplitude.
On the contrary, when the compensation is achieved
the density response to the applied modulations remains very weak even for long exposition times.
It is important to note that this prediction is valid for many-boson systems where the relative motion is not only 
in the ground but also in excited states. In particular, it means that the dynamical compensation
can work at non-zero temperatures as well.

\acknowledgments

Computation time on the bwGRiD and the Cray XE6 cluster ``Hermit'' at the HLRS in Stuttgart, and financial support
by the HGS MathComp, Minerva Short Term research grant, and the DFG also within the framework of the ``Enable fund'' of the
excellence initiative at Heidelberg university are greatly acknowledged.

\clearpage
\bibliography{calibration}

\newpage

\begin{table}
\begin{center}
\begin{tabular}{|c|c|c|c|}
\hline
M                    & N=10              & N=100             & N=1000 \\ \hline
1                    & 7.0\ul{71067811865483}& 70.\ul{71067811865483} & 707.\ul{1067811865483} \\ \hline
2                    & 7.038\ul{769026303168}& 70.6801\ul{6951747168} & 707.0764\ul{334257315} \\ \hline
3                    & 7.0383\ul{50652406389}& 70.680125\ul{41218675} & 707.07642898\ul{71865} \\ \hline
4                    & 7.0383484\ul{24909910}& 70.6801253917\ul{4549} & \\ \hline
5                    & 7.0383484153\ul{49058}& 70.680125391737\ul{62} &  \\ \hline
6                    & 7.038348415311\ul{494}&                        &  \\ \hline
7                    & 7.03834841531101\ul{8}&                        &  \\ \hline
\hline
$E_{\mathrm{exact}}$ & 7.038348415311011     & 70.68012539173752      & 707.0764289869851 \\ \hline
\end{tabular}
 \end{center}
\caption{Ground state energies of the harmonic interaction Hamiltonian for the systems of N=10,100,1000 bosons.
Exact analytical versus numerical MCTDHB($M$) results, $M$ is the number of self-consistent orbitals used.
The interparticle interaction strengths have been chosen to keep $\Lambda=K_0(N-1)=0.5$ constant.
In this case all these systems have the same Gross-Pitaevskii solution, i.e., the same energy per particle.
The one orbital MCTDHB($M=1$) theory is fully equivalent to the Gross-Pitaevskii mean-field.
It is seen that converged results are obtained with less self-consistent orbitals when increasing the number of particles.}
\label{tabHIMenergies}
\end{table}

\end{document}